# Progress in the Verification and Validation Efforts for START: A Spent Fuel Routing Tool

Harish R. Gadey, Kacey D. McGee, Patrick D. Royer

*Pacific Northwest National Laboratory, 902 Battelle Blvd, Richland, WA 99354, harish.gadey@pnnl.gov*

[leave space for DOI, which will be inserted by ANS]

## INTRODUCTION

The U.S. Department of Energy's (DOE) Office of Nuclear Energy (DOE-NE) is planning for an integrated waste management approach to transport, store, and eventually dispose of spent nuclear fuel (SNF) and other high-level radioactive waste (HLW) as part of the Integrated Waste Management (IWM) program [1]. In support of this effort, the Stakeholder Tool for Assessing Radioactive Transportation (START) is being developed within the IWM program [2]. This is a web-based decision support tool that can be used to analyze geospatial data related to the transportation of SNF and HLW.

START is designed as a web-based application using an ArcGIS server through which the user can select the origin and destination of the route [3]. This is followed by selecting the mode of transportation. Some of the modes available to the user include rail, heavy haul truck, and barge. The option of utilizing more than one mode of transportation (intermodal transportation) can also be selected in START. A few examples of intermodal transportation include barge to rail, and heavy haul truck to rail. Users can also select any stops or barriers they would like to introduce in the routes. This is followed by selection of the routing criteria. Three primary routing options available include minimum population, minimum distance, and minimum time. Apart from that, a few other options include accounting for a weighted average of the three aforementioned routing options. The next step involves the selection of the buffer distance of interest which includes the two choices available of 800 m and 2500 m, respectively. Finally, an option to select any prohibited rail carriers that the user does not wish to use is available. After making these selections, a route is ready to be created.

This work primarily focuses on the Verification & Validation (V&V) efforts currently being performed on the START tool. Several key outputs of interest are being tested including route population, route length, route population density, and incident free doses for the public as well as the crew.

## START OVERVIEW

Once a route is generated, key outputs from START include route length, travel time, buffer zone population, accident likelihood, incident free dose to the public, track class, etc. Apart from this, several other files can also be downloaded for further analysis including the shape, and keyhole markup language (KML) files. Fig. 1 shows an example of START routes from various locations across the United States.

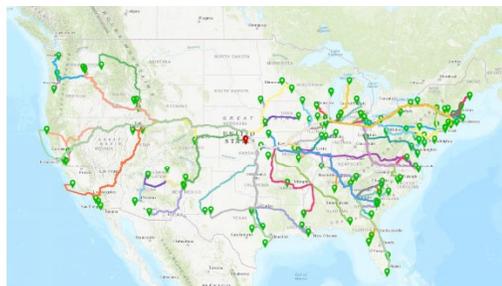

*Fig. 1. Sample START routes from various origin locations (green balloons) to Lebanon, KS, the geographic center of the continental United States (red balloon). NOTE: Routes are for illustrative and analytical purposes and do not represent planned SNF/HLW transport routes.*

Population data in the buffer zones are important outputs from the START tool. START uses LandScan population datasets which are updated yearly [4]. LandScan population datasets provide both day and night population data that is used in the tool.

DOE anticipates the user base of START could include a diverse range of organizations such as Federal, State, local, and Tribal government personnel. Nuclear utilities, transportation carriers, social scientists, researchers, students, etc. could be potential START users in the future. Therefore, START was designed to cater to the needs of diverse organizations with a key emphasis on providing flexibility.

In addition to generating routes and providing critical statistical information, the START user interface (UI) also hosts several base maps that the user can choose from including street networks, topographic maps, population density maps, and flood hazard layers. Spatial statistics can also be generated for a user defined area on the START UI. Once an area is defined (using a polygon, or a set distance or time from a given point), spatial statistics like number of fire stations, law enforcement agencies, hospitals, Tribal lands, schools etc. within the area are available to the user. The measurement tool on the START interface provides the user with options to measure area, distance, and identify latitude and longitude for a given point.

Routes in START can be created using the UI as discussed earlier and also using the batch mode option. This mode provides the user an option to input several routes at a given time and creates all the routes at once. Once the route is generated, the user is also provided an option to share or modify the routes. Apart from that, there is also an option to export route reports containing distance, population, accident frequency data, etc. on the START UI.

On top of performing routing and statistical analyses, START can also be used to perform training activities along the routes (proximity and response time studies for police, fire, and DOE's transportation emergency preparedness program personnel), inform and support communication between utilities and transportation infrastructure operators, and integrating with other DOE backend analysis tools such as the Next Generation System Analysis Model (NGSAM).

**AGILE SOFTWARE DEVELOPMENT**

Software development is a relatively complex process and requires the implementation of a strategy and methodology. Two of the most commonly used strategies for software development are the waterfall and agile software development process. The waterfall approach requires the different work teams to complete a given phase or set of tasks before moving on to the next phase. On the other hand, an agile methodology encourages following an iterative process with regular interactions between the development and the V&V teams. The agile software development methodology was decided to be used for START development due to enhanced communication and feedback opportunities between the teams. The V&V team tests the features implemented in the tool. Any suggestions provided by the V&V team are implemented by the development team. These features are then subsequently tested by the V&V team and any feedback is shared with the development team. Fig. 2 shows an example of the agile software development process.

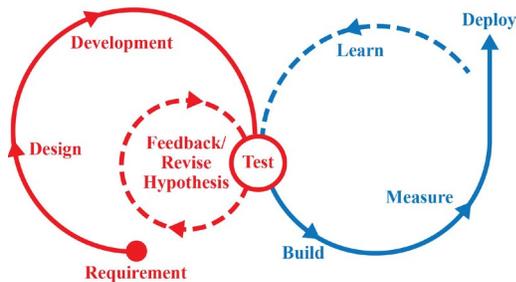

*Fig. 2. An example of an agile software development life cycle.*

The latest version of START (version 3.3) was released in August 2022. The major update in this version of START is migrating from the Idaho National Laboratory servers over to the Amazon Web Services (AWS) servers. This latest version also includes using updated geographic information systems (GIS) data layers.

**START V&V EFFORTS**

The START V&V efforts were initially started in 2020. These efforts continued during 2021 and 2022, focusing primarily on the subsequent releases of START [5 – 8]. Initially a total of 149 hypothetical origin-destination pairs were established to test START. The criteria used for creating the test routes include:
- Using minimum distance routing algorithm for all routes.
- Creating routes using both buffer zones (800 m and 2500 m).
- Choosing the geographical center of the continental United States as the destination.
- Choosing origin sites with operating, shutdown, and research reactors. Also, facilities where DOE owned SNF and/or HLW may be stored were also chosen.

The primary aim for the V&V testing effort in 2023 was the expansion of the test series. The expanded routes included the same origin-destination pairs that were previously used but involved the use of variable modes of transportation for some of the routes. This was informed by the mode of transportation suggested in a DOE study, as well as the modes used in the NGSAM tool [9, 10]. This expanded test series consists of a total of 169 unique route configurations. In these V&V efforts the following key outputs were tested:
- Total route length
- Population along the route length at buffer distance of 800 m and 2500 m, respectively
- Population density at buffer distance of 800 m and 2500 m, respectively
- Incident free dose to the public and the transportation crew.

**Total Route Length**

Route length is one of the key outputs provided by START. A route that is generated in START consists of several individual segments whose distance is summed to obtain a total length of the route. Out of a total of 169 unique route configurations, 150 routes ran successfully. 16 origin-destination pairs failed to generate a route, and for three of the routes the transportation mode of interest was not available in START for the origin sites.

The route length was extracted for all the routes that ran successfully. The shapefiles were processed using the Quantum Geographic Information System (QGIS), an open-source GIS tool to compare the START provided route lengths and the QGIS provided route lengths. The minimum and maximum differences in route distance reported in START compared to the independent QGIS analysis was found to be 0.00 and 14.39%, respectively. The average difference between the distance reported from START and QGIS for all 150 origin-destination pairs was determined to be 0.47%.

**Route Population Analysis**

Population is one of the important outputs that is obtained from START. In the START UI, the population along a route is generally obtained by calculating buffer statistics. The individual segments in the route are concatenated to form a single geometry from the origin to a destination. The buffer zones were subsequently created using this concatenated geometry. A 1:2 weighted average of the day and nighttime LandScan population data was utilized to estimate the total population in a given buffer zone along a route of interest.

Out of a total of 300 routes (both buffer zones), population statistics were obtained for 124 of the routes. Out of the 124 routes that produced buffer zone population statistics, 53 were obtained from routes with a 2500 m buffer zone and the remaining 71 results were estimated using routes with an 800 m buffer zone. Using the aforementioned methodology, a minimum, maximum, and average population difference in both buffer zones was calculated to be 0.01, 38.75, and 4.95%, respectively. A population difference of greater than 10% was recorded in 10 instances out of the 124 routes that produced the buffer zone population results.

**Route Population Density Analysis**

Population density captures the ratio of the number of people in a given buffer zone to the area of the buffer zone. In previous versions of the V&V efforts, population density was calculated at a segment level and summed, however, this approach presented a challenge of double counting the population in various segments. Therefore, all the individual segments were concatenated to form a single geometry and the 800 m and 2500 m buffer zones were established using the single route geometry.

The population density data from START was obtained by generating buffer statistics for the entire route. The minimum, maximum, and average difference between the START data and the QGIS generated data was found to be 0.08, 38.19, and 4.88%, respectively. Ten of the 124 routes exhibited a difference in population density exceeding 10%.

**Incident Free Dose Analysis**

In the current version of START (version 3.3), the dose values are reported at an individual segment level. The dose calculations were performed by incorporating the concept of unit dose factors that was introduced in 2015. This concept assumes that the dose rate of the packages is at the current US regulatory limit of 1mSv/h at a distance of 2 m. It is also worth noting that the unit dose methodology is currently only applicable to routes with an 800 m buffer; the doses are not applicable to routes with a 2500 m buffer.

*Dose Analysis for Waterway Shipments*

A segment level analysis of all the routes that employ barge mode of transportation was performed in this work. It was identified that over 110 barge segments exist in the 150 routes that ran successfully. A maximum difference of 0.21 and 0.02% was obtained for public and crew doses, respectively. The maximum difference in START estimated and independent calculated values for both incident free public and crew dose was found to be $1.25 \times 10^{-11}$ mSv.

*Dose Analysis for Rail Shipments*

A segment analysis of all the routes using the rail mode of transportation was performed in this study as well. This included a total of over 79,000 individual rail segments. A maximum difference in dose between START reported values and independent calculations was obtained to be 13.03 and 4.0% for the public and crew dose, respectively. In absolute values, this difference translates to $2.58 \times 10^{-9}$ and $1.40 \times 10^{-11}$ mSv for the public and crew dose, respectively.

*Dose Analysis for Highway Shipments*

Out of all the test routes that were successfully created in START, there are over 339,000 individual segments that were analyzed in this work. Highway doses for the public were reported for both on-link and off-link population exposures. The maximum crew, on-link, and off-link difference in dose was found to be $1.63 \times 10^{-11}$, $1.01 \times 10^{-8}$, and $1.31 \times 10^{-8}$ mSv, respectively.

**Route Aberrations**

Several qualitative studies were also carried out to gain an insight of the performance of START. These include closely observing the created routes in START for any abnormalities such as deviation of the route through non-traversable networks or areas, differences between the shape and KML files, non-required transload operations etc. Fig. 3 provides a few examples of unexpected routing behavior. Once this behavior is identified, the START development team is notified and work is initiated to investigate and correct the cause of these aberrations. Common reasons could include geospatial data, metadata, or routing algorithm configuration errors.

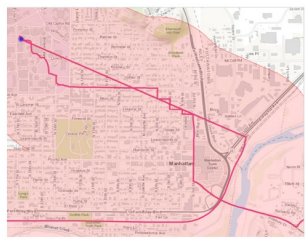
Route Jumps Across Streets

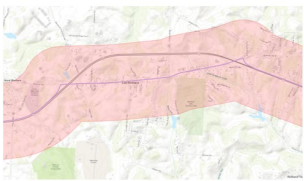
Detour From Highway

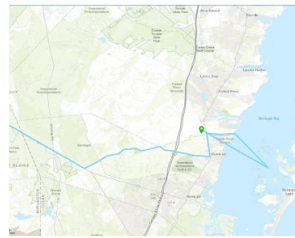
Route Movement Across Water

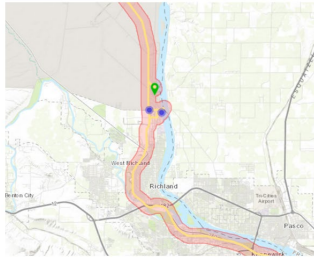

Fig. 3. Qualitative observations for START routes.

## CONCLUSION

This paper initially discusses the various functionalities and features available in the tool to meet requirements of both decision makers and stakeholders. The V&V efforts performed in the latest version of START (version 3.3) were then explored. These efforts found good agreement with the analysis performed using QGIS for the total route distance, route population in the buffer zones, and the population density. The difference in total route length was found to be <0.5% on average and the difference in average population metrics such as total route population and population density were found to be <5%. The independent dose assessments found reasonably good agreement with the data reported in START. Apart from that, some high-level routing aberrations and behavior were also explored in the work. The START development and the START V&V teams continue to collaborate and identify areas for further improvement of START.

## ENDNOTES

This is a technical presentation that does not take into account contractual limitations or obligations under the Standard Contract for Disposal of Spent Nuclear Fuel and/or High-Level Radioactive Waste (Standard Contract) (10 CFR Part 961). To the extent discussions or recommendations in this presentation conflict with the provisions of the Standard Contract, the Standard Contract governs the obligations of the parties, and this report in no manner supersedes, overrides, or amends the Standard Contract. This presentation reflects technical work which could support future decision making by DOE. No inferences should be drawn from this presentation regarding future actions by DOE, which are limited both by the terms of the Standard Contract and Congressional appropriations for the Department to fulfill its obligations under the Nuclear Waste Policy Act including licensing and construction of a spent nuclear fuel repository.

This information was prepared as an account of work sponsored by an agency of the U.S. Government. Neither the U.S. Government nor any agency thereof, nor any of their employees, makes any warranty, expressed or implied, or assumes any legal liability or responsibility for the accuracy, completeness, or usefulness, of any information, apparatus, product, or process disclosed, or represents that its use would not infringe privately owned rights. References herein to any specific commercial product, process, or service by trade name, trademark, manufacturer, or otherwise, does not necessarily constitute or imply its endorsement, recommendation, or favoring by the U.S. Government or any agency thereof. The views and opinions of authors expressed herein do not necessarily state or reflect those of the U.S. Government or any agency thereof.

## ACKNOWLEDGEMENT

Pacific Northwest National Laboratory (PNNL) is operated by Battelle Memorial Institute for the U.S. Department of Energy (DOE) under Contract No. DE-AC05-76RL01830. This work was supported by the DOE Office of Integrated Waste Management.